\documentclass{article}
\usepackage[T1]{fontenc}
\usepackage[utf8]{inputenc}
\usepackage[english]{babel}
\usepackage{lmodern}
\usepackage{microtype}
\usepackage{amsmath,amssymb}
\usepackage{graphicx}
\usepackage{caption}
\usepackage{subcaption}
\usepackage{booktabs}
\usepackage{siunitx}
\usepackage{authblk}
\usepackage{enumitem}
\usepackage{float}
\usepackage[section]{placeins}
\usepackage{tikz}
\usetikzlibrary{arrows.meta,positioning,fit,backgrounds}
\usepackage[unicode]{hyperref}
\title{Predictive Criteria for Electrospray-Assisted Droplet Dynamics in Aerodynamic Flow Fields}

\author[1]{Ethan Kahn\thanks{Email: \texttt{ethankahn@protonmail.com}}}
\affil[1]{Unaffiliated, Zürich, Switzerland}
\affil[ ]{\texttt{ORCID: \href{https://orcid.org/0009-0003-4156-8356}{0009-0003-4156-8356}}}

\date{August 2025}

\begin{document}

\maketitle

\section*{Abstract}
Electrospray technology enables external electric fields to steer charged droplets, with potential applications in fuel–air mixing and aerodynamic flow control. This study develops a computational framework that couples a steady OpenFOAM RANS solution of a NACA 1912 airfoil channel flow with a reduced-order Lagrangian particle model to examine droplet trajectories under viscous drag and electrostatic forces. Baseline uniform-flow tests confirmed that electrostatic deflection weakens as inertial effects grow with freestream velocity. In the non-uniform aerodynamic field, an opposing streamwise electric field increased residence times and produced local reversal in low-speed pockets. Building on these results, we derive a predictive criterion linking the minimum electric field for reversal to the local convective velocity and introduce a “control authority” map that highlights regions where modest fields achieve strong kinematic response. Together, these diagnostics provide a design-oriented basis for positioning electrodes and tuning field strength in aerodynamic environments. The framework thus establishes both a computationally efficient tool for parametric studies and predictive criteria for electrospray-assisted flow manipulation, laying groundwork for three-dimensional extensions and experimental validation.

\noindent\textbf{Keywords: Electrospray; Electrohydrodynamics; Droplet dynamics; Flow control; Predictive criteria; Airfoil aerodynamics} 

\section{Introduction}

Electrospray technology harnesses strong electric fields to generate and steer highly charged microdroplets with unusual precision, enabling advances in propulsion, analytical chemistry, and materials processing. In astronautics, colloid/electrospray thrusters deliver micronewton-class thrust with high electrical-to-kinetic efficiency and increasingly compact, microfabricated architectures, while emitter geometry continues to be optimized to limit plume divergence and improve performance \cite{ramos-tomas_impact_2024,cisquella-serra_scalable_2022,grustan-gutierrez_microfabricated_2017,gamero-castano_electrospray_2000,gamero-castano_characterization_2004}. In analytical chemistry, electrospray ionization (ESI) transformed mass spectrometry by extending sensitive analysis to large biomolecules; subsequent developments refined source physics and quantitative methodology \cite{yamashita_electrospray_1984,fenn_electrospray_1989,wilm_analytical_1996,kebarle_electrospray_2009,banerjee_electrospray_2012,fernandez_de_la_mora_ever_2019,makarov_fundamental_2017}. Beyond diagnostics and propulsion, electrosprays enable thin-film deposition and nanoparticle fabrication, including controlled, monodisperse particles and coatings on complex 3D surfaces \cite{jaworek_classification_1999,barrero_micro-_2007,berkland_fabrication_2001,kim_electrospray_2024,jaworek_review_2018}. More recently, electrohydrodynamic (EHD) effects have been leveraged to manipulate combustible sprays, with external fields used to stabilize dispersion, extend droplet–gas interaction times, and enhance fuel–air mixing \cite{gomez_charge_1994,shrimpton_characterisation_2001,kim_enhancement_2011,fredrich_electrostatic_2023,liu_review_2017,wang_analysis_2013,nkoi_comparative_2015}. Together, these threads illustrate the cross-disciplinary relevance of electrosprays to aerospace and energy applications.

At the physics level, steady cone–jet emission arises from a balance of electrostatic stress, capillarity, and inertia. Foundational analyses established the deformation and breakup of electrified drops \cite{taylor_disintegration_1964} and produced scaling laws that organize the cone–jet regime and electrified disintegration across fluids and operating conditions \cite{ganan-calvo_cone-jet_1997,collins_universal_2013,ganan-calvo_onset_2016}. In propulsion-relevant operation, ionic liquids offer broad stability windows but challenge models due to high conductivity, viscosity, and charge-relaxation timescales; multi-emitter arrays additionally introduce concerns such as off-axis plume broadening and thrust-vector misalignment \cite{uchizono_emission_2020,cisquella-serra_scalable_2022,ramos-tomas_impact_2024}. These complexities reinforce the need for modeling approaches that bridge nanoscale interfacial physics and device-scale plume evolution.

Computationally, electrospray is an intrinsically multi-scale, multi-physics problem. Interfacial charge separation, jet initiation, and early breakup occur at submicron scales, while downstream plume transport, space-charge evolution, and evaporation unfold over millimeters to meters. Recent work employs hybrid strategies that couple molecular or mesoscopic descriptions of the near field with continuum or particle-based solvers for the plume, including particle-in-cell and reduced-order transport models \cite{asher_multi-scale_2022,johnson_computational_2024}. Such studies show that space-charge can feed back on emission characteristics, while solvent evaporation continuously perturbs conductivity and surface tension in flight, reshaping droplet stability and charge partitioning \cite{johnson_computational_2024}. New scaling and modeling efforts further refine the description of ion-emission regimes and characteristic lengths pertinent to cone–jet operation \cite{perez-lorenzo_modelling_2022,magnani_determination_2025}. Despite these advances, most experimental and computational studies emphasize quiescent or vacuum environments (e.g., space thrusters or atmospheric-pressure ESI without net crossflow). By contrast, the dynamics of charged droplets embedded in fast, spatially non-uniform aerodynamic flows remain comparatively underexplored, even though prior EHD-spray and combustion studies indicate strong potential for field-mediated stabilization and residence-time control \cite{gomez_charge_1994,shrimpton_characterisation_2001,kim_enhancement_2011,fredrich_electrostatic_2023}.

The present study addresses this gap by examining electrospray-inspired droplet dynamics in a high-speed, non-uniform aerodynamic environment representative of an airfoil flow. Building on recent demonstrations of EHD-assisted spray control in combusting and non-combusting contexts \cite{fredrich_electrostatic_2023}, we develop a hybrid framework that couples a high-fidelity RANS solution around a NACA~1912 airfoil (OpenFOAM) with a lightweight Lagrangian particle model for charged droplets. The formulation isolates the essential drag–electric forcing physics while remaining computationally efficient for parametric sweeps. Specifically, we interrogate whether externally applied electric fields can (i) counteract convective transport, (ii) extend droplet residence times within low-speed regions created by the aerodynamic field, and (iii) locally reverse streamwise motion near surfaces. By bridging electrospray physics with aerodynamic flow simulation, this work provides a controlled testbed for EHD flow-control concepts and lays the groundwork for three-dimensional extensions, incorporation of evaporation and space-charge interactions, and experimental validation in well characterized aerodynamic conditions \cite{cisquella-serra_scalable_2022,asher_multi-scale_2022,johnson_computational_2024}.

\section{Materials and Methods}\label{sec:methods}

\subsection{Overview}
The simulation framework combines a steady-state computational fluid dynamics (CFD) solution with a lightweight Lagrangian electrospray droplet model. The airflow is first resolved using OpenFOAM~v12 under Windows Subsystem for Linux (WSL, Windows~11, 12-core AMD Ryzen CPU). The converged velocity and pressure fields provide a frozen background for subsequent particle integration. One-way coupling (fluid $\rightarrow$ particle) is assumed, appropriate for dilute sprays where feedback on the gas phase and applied electric field is negligible.

\subsection{Computational domain and geometry}
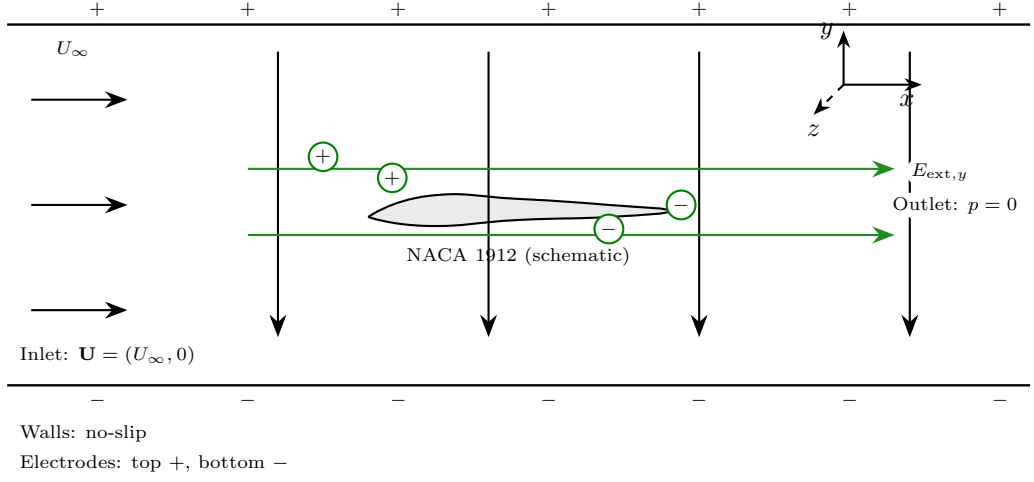
\begin{figure}[H]
\centering
\begin{tikzpicture}[x=0.4cm,y=0.4cm,>=Stealth,thick,
    charge/.style={font=\scriptsize},
    droplet/.style={circle,inner sep=1pt,draw=green!50!black,thick,font=\scriptsize},
    labelbox/.style={fill=white,rounded corners,inner sep=1.5pt,outer sep=1pt},
    every node/.style={align=left}]
\definecolor{fieldgreen}{RGB}{34,139,34}

\draw[line width=0.9pt] (0,0) -- (34,0);   
\draw[line width=0.9pt] (0,12) -- (34,12); 

\foreach \x in {3,8,13,18,23,28,33}{
  \node[charge] at (\x,12.5) {$+$};
  \node[charge] at (\x,-0.5) {$-$};
}

\foreach \y in {2.5,6,9.5}{
  \draw[-{Stealth[length=3mm]}] (0.8,\y) -- (4.0,\y);
}

\node[labelbox,font=\scriptsize] at (2.2,11.2) {$U_\infty$};

\begin{scope}
  \path[fill=black!8,draw=black]
    plot[smooth,tension=0.9] coordinates {(12,5.6) (14,6.3) (17,6.2) (20,6.0) (22,5.8)
                                          (20,5.6) (17,5.5) (14,5.3) (12,5.6)};
\end{scope}
\node[font=\scriptsize,labelbox] at (17,4.3) {NACA 1912 (schematic)};

\foreach \x in {9,16,23,30}{
  \draw[-{Stealth[length=3mm]}] (\x,11.1) -- (\x,1.6);
}
\node[labelbox,font=\scriptsize] at (31.0,7.1) {$E_{\mathrm{ext},y}$};

\foreach \y in {5.0,7.2}{
  \draw[fieldgreen, -{Stealth[length=3mm]}] (8,\y) -- (29.5,\y);
}
\node[fieldgreen,labelbox,font=\scriptsize] at (31.0,5.9) {$E_{\mathrm{ext},x}$};

\node[droplet,fill=white] at (10.5,7.6) {$+$};
\node[droplet,fill=white] at (12.8,6.9) {$+$};
\node[droplet,fill=white] at (20.0,5.2) {$-$};
\node[droplet,fill=white] at (22.4,6.0) {$-$};

\node[anchor=west,font=\scriptsize,labelbox] at (0.2,1.0) {Inlet: $\mathbf U=(U_\infty,0)$};
\node[anchor=east,font=\scriptsize,labelbox] at (33.8,6.0) {Outlet: $p=0$};
\node[anchor=west,font=\scriptsize,labelbox] at (0.2,-1.6) {Walls: no-slip};
\node[anchor=west,font=\scriptsize,labelbox] at (0.2,-2.6) {Electrodes: top $+$, bottom $-$};

\coordinate (axorig) at (27.8,10.0);
\draw[->] (axorig) -- ++(2.6,0) node[below left,xshift=1pt] {$x$};
\draw[->] (axorig) -- ++(0,1.8) node[left] {$y$};
\draw[->,dashed] (axorig) -- ++(-1.0,-1.0) node[below] {$z$};

\end{tikzpicture}
\caption{Schematic of the computational domain and boundary conditions. A NACA~1912 profile is centered in a rectangular channel with inlet velocity $\mathbf{U}=(U_\infty,0,0)$, fixed-pressure outlet $p=0$, and no-slip walls. Oppositely charged plates generate vertical $E_{\mathrm{ext},y}$ and streamwise $E_{\mathrm{ext},x}$ components. The coordinate system follows $x$ (streamwise), $y$ (wall-normal), and $z$ (spanwise).}
\label{fig:geometry}
\end{figure}

The test case consists of a NACA~1912 airfoil of chord $c=\SI{0.2}{m}$ and span $L_s=\SI{0.2}{m}$ placed at the center of a rectangular channel of length $L=\SI{1.0}{m}$, height $H=\SI{0.2}{m}$, and width $W=\SI{0.2}{m}$. The airfoil is aligned with the streamwise $x$-axis at angle of attack $\alpha=\SI{0}{\degree}$. The coordinate system follows OpenFOAM convention: $x$ streamwise, $y$ wall-normal, and $z$ spanwise. The resulting chord Reynolds number is
\[
Re_c = \frac{\rho U_\infty c}{\mu} \approx 1.3\times 10^5,
\]
for freestream velocity $U_\infty=\SI{10}{m\,s^{-1}}$, air density $\rho=\SI{1.18}{kg\,m^{-3}}$, and dynamic viscosity $\mu=\SI{1.85e-5}{Pa\,s}$. Blockage ratio is approximately $t/H\approx 0.12$ based on the 12\% thick profile. For clarity, Figure~\ref{fig:geometry} sketches the computational domain (rectangular channel with a centered NACA~1912 airfoil), the inlet/outlet and no-slip wall boundary conditions, the electrode polarities generating $E_{\mathrm{ext},x}$ and $E_{\mathrm{ext},y}$, and the $(x,y,z)$ coordinate directions.

\subsection{Droplet dynamics and numerics}\label{sec:droplet}
Each droplet is treated as a rigid, charged sphere of radius $r_d$ and density $\rho_d$ moving in a prescribed CFD background field $\mathbf{u}_f(\mathbf{x})$ with dynamic viscosity $\mu$. One-way coupling (fluid $\to$ particle) is assumed, appropriate for dilute sprays where feedback on the gas phase and applied field is negligible.

\paragraph{Equations of motion.}
The Lagrangian droplet evolves according to
\begin{align}
\dot{\mathbf{x}}_d &= \mathbf{v}_d, \\
m_d\,\dot{\mathbf{v}}_d &= 6\pi\mu r_d\big(\mathbf{u}_f(\mathbf{x}_d)-\mathbf{v}_d\big) + q_d\,\mathbf{E}_{\text{ext}},
\end{align}
with $m_d=\tfrac{4}{3}\pi r_d^3\rho_d$ and a uniform applied field $\mathbf{E}_{\text{ext}}=(E_{\text{ext},x},E_{\text{ext},y},0)$. The particle Reynolds number $\displaystyle Re_d=\frac{2\rho r_d \lVert \mathbf{u}_f-\mathbf{v}_d\rVert}{\mu}\ll1$, so Stokes drag is appropriate; standard $O(Re_d^{2/3})$ corrections were neglected in production runs.

It is convenient to define the droplet relaxation time and electric acceleration
\begin{equation}
\tau_d=\frac{m_d}{6\pi\mu r_d}=\frac{2\rho_d r_d^2}{9\mu}, 
\qquad 
\mathbf{a}_E=\frac{q_d}{m_d}\,\mathbf{E}_{\text{ext}},
\end{equation}
so that
\begin{equation}
\dot{\mathbf{v}}_d=\frac{\mathbf{u}_f(\mathbf{x}_d)-\mathbf{v}_d}{\tau_d}+\mathbf{a}_E.
\end{equation}
Two convenient nondimensional parameters used below are the Stokes number $\displaystyle \text{St}=\frac{\tau_d U_\infty}{c}$ and the electric-to-convective ratio
\begin{equation}
\Lambda_E=\frac{\lVert q_d\mathbf{E}_{\text{ext}}\rVert}{6\pi\mu r_d U_\infty}
=\frac{\lVert \mathbf{a}_E\rVert\,\tau_d}{U_\infty}.
\end{equation}

\paragraph{Time integration and interpolation.}
Trajectories are advanced with explicit Euler,
\begin{align}
\mathbf{v}_d^{n+1} &= \mathbf{v}_d^{n} + \Delta t\left[ \frac{\mathbf{u}_f(\mathbf{x}_d^{n})-\mathbf{v}_d^{n}}{\tau_d} + \mathbf{a}_E \right],\\
\mathbf{x}_d^{n+1} &= \mathbf{x}_d^{n} + \Delta t\,\mathbf{v}_d^{n},
\end{align}
using $\Delta t\sim10^{-6}\,$s. The background velocity is sampled on the mid-span structured slice and evaluated at particle positions by tri-linear interpolation.

\paragraph{Termination and timestep constraints.}
A trajectory terminates upon (i) wall or airfoil impingement, (ii) domain exit, or (iii) reaching $t>10\,$ms. For robustness we enforce $\Delta t \le 0.1\,\tau_d$ and a convective constraint $\displaystyle \Delta t \le 0.25\,\Delta_{\text{grid}}/\lVert\mathbf{u}_f\rVert_\infty$.

\paragraph{Scope of the model.}
Lift, added-mass, and history terms are omitted, consistent with $Re_d\ll1$ and $\rho/\rho_d\ll1$ over the short times considered. Evaporation, Coulomb fission, and space-charge interactions are neglected here and identified as future extensions.

\paragraph{Computational cost.}
With the frozen-field assumption and explicit timestepping, a single trajectory executes in under a second on one CPU core, enabling sweeps over $(r_d,q_d,\mathbf{E}_{\text{ext}})$.

\subsection{Meshing strategy}
\begin{figure}[h]
  \centering
  \includegraphics[width=0.48\linewidth]{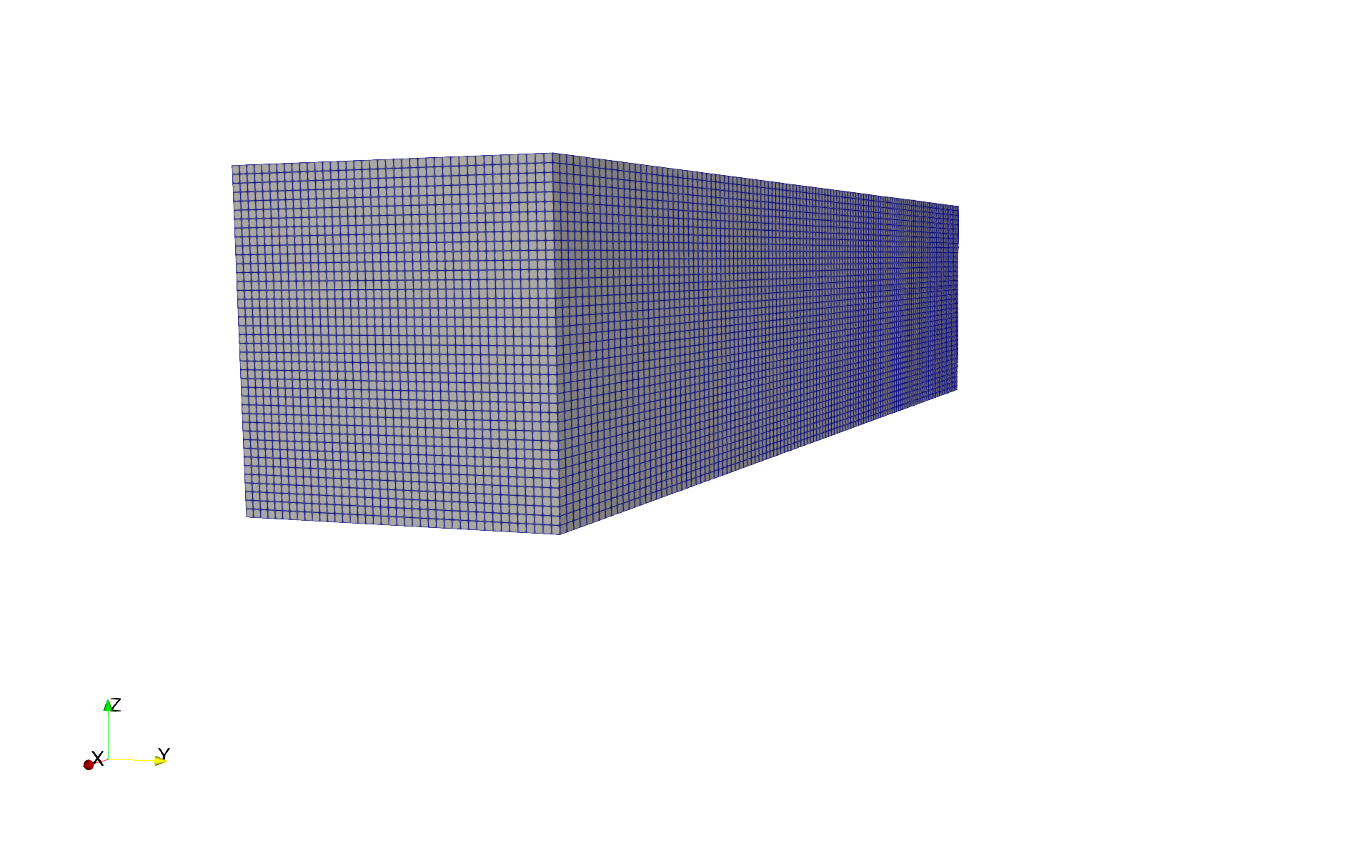}
  \includegraphics[width=0.48\linewidth]{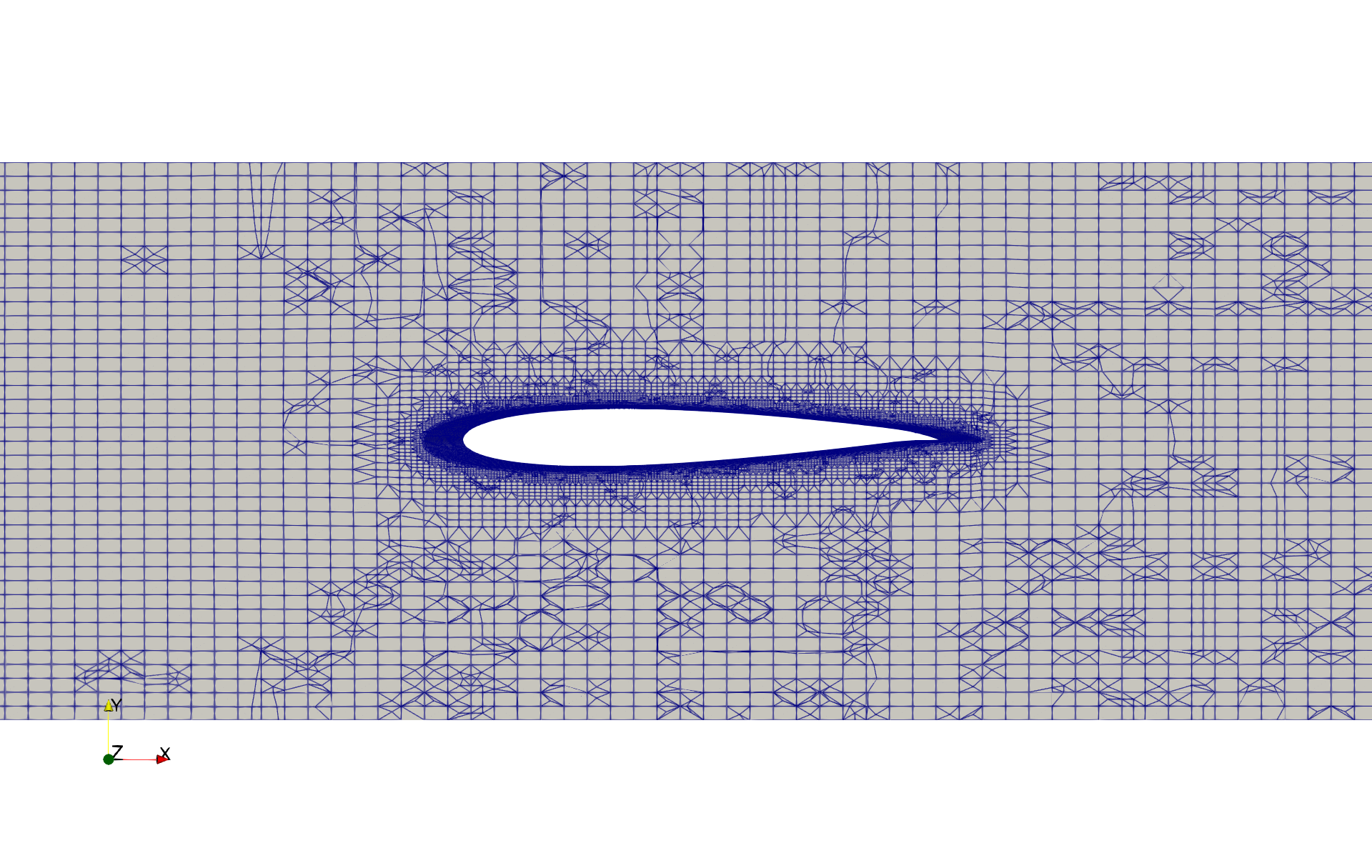}
  \caption{Computational mesh of the NACA~1912 airfoil in a confined channel
($L=\SI{1.0}{m}$, $H=W=\SI{0.2}{m}$). (a) Three-dimensional view of the domain showing inlet and side walls. (b) Close-up of the refinement region around the airfoil surface. The mesh contains $1.59\times 10^6$ predominantly hexahedral cells with minimum non-orthogonality $<20^\circ$.}
  \label{fig:mesh}
\end{figure}

The geometry was provided as an STL file (see Supplementary Material, Geometry~S1) and imported via \texttt{constant/triSurface}. Mesh generation proceeded in two steps:
\begin{enumerate}[nosep,leftmargin=*]
  \item A structured base grid was created using \texttt{blockMesh}.
  \item Local refinement around the airfoil was applied with \texttt{snappyHexMesh}, including castellation, surface snapping, and optional prism layer addition.
\end{enumerate}
The final mesh comprised $1.59\times10^6$ predominantly hexahedral cells with minimum non-orthogonality below $20^\circ$. No custom wall-layer settings were applied, consistent with $k$--$\varepsilon$ wall-function treatment. Mesh quality statistics from \texttt{checkMesh} are summarized in Table~\ref{tab:mesh}. A representative mesh view is shown in Figure~\ref{fig:mesh}.

\begin{table}[h]
\centering
\caption{Mesh quality statistics (from \texttt{checkMesh}).}
\label{tab:mesh}
\begin{tabular}{ll}
\toprule
Total cells & $1.59\times 10^6$ \\
Base cell size $\Delta_{\mathrm{base}}$ & $\sim\SI{3}{mm}$ \\
Max non-orthogonality & $<20^\circ$ \\
Max skewness & $<4$ \\
Max aspect ratio & $<10$ \\
\bottomrule
\end{tabular}
\end{table}

\subsection{Boundary and initial conditions}
Boundary conditions are summarized in Table~\ref{tab:bc}. Turbulence quantities at the inlet were prescribed using intensity $I=5\%$ and length scale $L_t=0.07H$, corresponding to $k=\SI{0.375}{m^2\,s^{-2}}$ and $\epsilon=\SI{2.7}{m^2\,s^{-3}}$ with $C_\mu=0.09$. Wall boundaries employed standard wall functions for $k$ and $\epsilon$. 

\begin{table}[h]
\centering
\caption{Boundary conditions for the CFD simulation.}
\label{tab:bc}
\begin{tabular}{llll}
\toprule
Patch & $U$ & $p$ & Turbulence \\
\midrule
Inlet  & fixedValue $(U_\infty,0,0)$ & zeroGradient & fixedValue $k,\epsilon$ \\
Outlet & zeroGradient                & fixedValue $p=0$ & zeroGradient \\
Airfoil, walls & noSlip              & zeroGradient & wall functions \\
Front/Back (spanwise) & noSlip       & zeroGradient & wall functions \\
\bottomrule
\end{tabular}
\end{table}

\subsection{Solver settings and convergence}
Simulations were performed with \texttt{simpleFoam}, OpenFOAM’s steady-state incompressible RANS solver. Discretization used second-order linear schemes for gradients and upwind/linearUpwind for convection. Pressure was solved with \texttt{GAMG} and velocity/turbulence with \texttt{smoothSolver} (Gauss–Seidel). Residuals for $U$, $p$, $k$, and $\epsilon$ were required below $10^{-4}$. Lift and drag coefficients were monitored until variations were under 0.1\% over 500 iterations. Simulations converged in $\sim2000$ iterations on 12 CPU cores.

\subsection{Workflow summary}
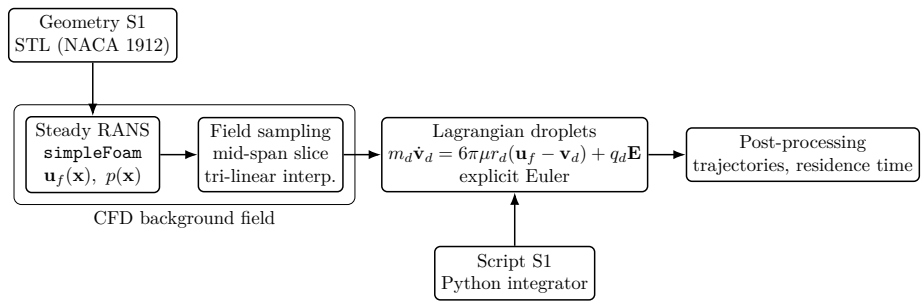
\begin{figure}[h]
  \centering
  \resizebox{\textwidth}{!}{%
    \begin{tikzpicture}[
        node distance=7mm,
        box/.style={draw, rounded corners=3pt, align=center, minimum width=25mm, minimum height=10mm, thick},
        arrow/.style={-Latex, thick}
      ]

      \node[box] (cfd) {Steady RANS \\ \texttt{simpleFoam} \\ $\mathbf u_f(\mathbf x),\ p(\mathbf x)$};
      \node[box, right=of cfd] (sample) {Field sampling \\ mid-span slice \\ tri-linear interp.};
      \node[box, right=of sample] (lag) {Lagrangian droplets \\ $m_d \dot{\mathbf v}_d = 6\pi\mu r_d(\mathbf u_f-\mathbf v_d)+q_d\mathbf E$ \\ explicit Euler};
      \node[box, right=of lag] (post) {Post-processing \\ trajectories, residence time};

      \draw[arrow] (cfd) -- (sample);
      \draw[arrow] (sample) -- (lag);
      \draw[arrow] (lag) -- (post);

      \node[box, above=10mm of cfd] (geo) {Geometry~S1 \\ STL (NACA 1912)};
      \draw[arrow] (geo) -- (cfd);

      \node[box, below=10mm of lag] (script) {Script~S1 \\ Python integrator};
      \draw[arrow] (script.north) -- (lag.south);

      \begin{scope}[on background layer]
        \node[draw, rounded corners=4pt, fit=(cfd)(sample), inner sep=6pt,
              label={[align=center]south:CFD background field}] {};
      \end{scope}

    \end{tikzpicture}%
  }
  \caption{Numerical workflow: CFD solution provides $\mathbf u_f(\mathbf x)$ and $p(\mathbf x)$; a mid-span slice is sampled for tri-linear lookups by the Lagrangian droplet integrator; post-processing yields trajectory and residence-time diagnostics.}
  \label{fig:workflow}
\end{figure}

The computational workflow (Figure~\ref{fig:workflow}) consisted of: (i) steady RANS simulation with \texttt{simpleFoam}, (ii) sampling of the mid-span velocity field, (iii) Lagrangian droplet integration under drag and electric forces, and (iv) post-processing to compute trajectory statistics. This pipeline isolates the essential physics of electrospray–flow interaction while remaining computationally efficient.

\section{Results}\label{sec:results}

This section reports (i) the baseline aerodynamics from the steady RANS solution at the mid–span plane ($z=0$), (ii) droplet trajectories in \emph{uniform} flows under a \emph{vertical} electric field, and (iii) droplet trajectories in the \emph{spatially varying} CFD field under a \emph{streamwise} electric field. In all cases, the electric field $\mathbf E$ acts only on the Lagrangian droplets; the CFD background field is unaffected.

\subsection{Baseline CFD flow field at the mid–span plane}\label{sec:results_cfd}

\begin{figure}[t]
  \centering
  \includegraphics[width=0.9\linewidth]{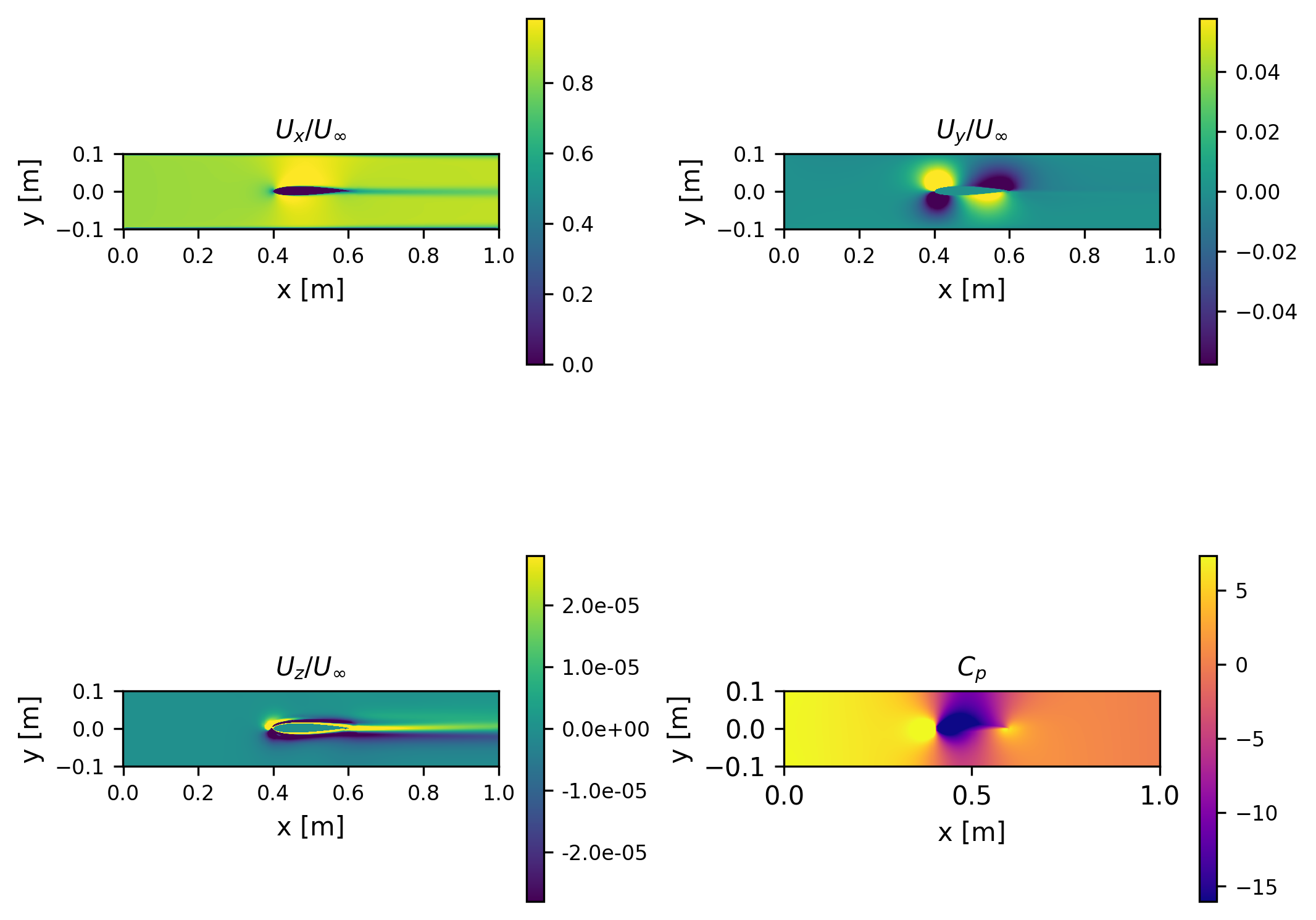}
  \caption{Baseline CFD solution on the mid–span plane ($z=0$).
  Panels show (a) streamwise velocity $U_x/U_\infty$, 
  (b) wall–normal velocity $U_y/U_\infty$, 
  (c) spanwise velocity $U_z/U_\infty$, 
  and (d) pressure coefficient $C_p$.
  The three velocity components are plotted with the same colormap for consistency, 
  but each uses its own scale (note the different ranges) so that weaker components 
  such as $U_y$ and $U_z$ remain visible. The aspect ratio is preserved to match the geometry.}
  \label{fig:cfd4}
\end{figure}

Figure~\ref{fig:cfd4} summarizes the baseline mid–span solution: $U_x$, $U_y$, $U_z$, and $C_p$ from top to bottom. Quantitative ranges extracted from the sampled field (file \texttt{airfoildataz0.csv}) are summarized below to anchor the qualitative views:
\begin{itemize}[nosep,leftmargin=*]
  \item Streamwise velocity: $U_x \in [0,\,\SI{12.5}{m\,s^{-1}}]$, median $\SI{10.0}{m\,s^{-1}}$; local acceleration up to $1.25\,U_\infty$.
  \item Vertical velocity: $U_y \in [-\SI{5.36}{m\,s^{-1}},\,\SI{5.58}{m\,s^{-1}}]$, i.e., $\lvert U_y\rvert \le 0.56\,U_\infty$.
  \item Spanwise velocity: $\lvert U_z\rvert \le \SI{0.048}{m\,s^{-1}} \approx 0.005\,U_\infty$ (negligible on this slice).
  \item Static pressure coefficient: $C_p \in [-0.25,\,0.65]$ with a suction minimum above the leading edge and a stagnation maximum at the nose.
\end{itemize}

\FloatBarrier

\subsection{Uniform–flow droplet trajectories under a vertical electric field}\label{sec:results_uniform}

\begin{figure}[t]
  \centering
  \includegraphics[width=\linewidth]{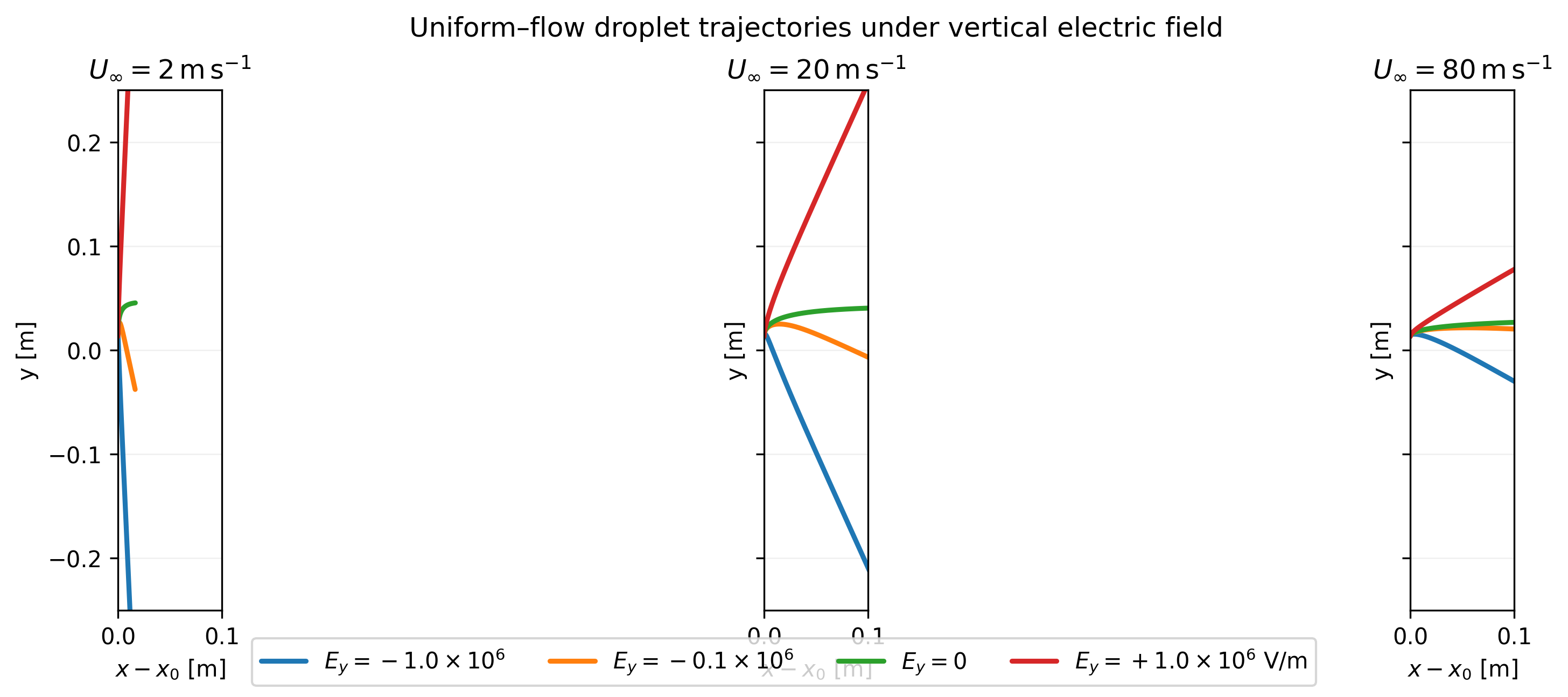}
  \caption{\textbf{Uniform–flow droplet trajectories under a vertical electric field.}
  Three panels compare freestream velocities $U_\infty=\{2,20,80\}\,\mathrm{m\,s^{-1}}$, 
  with trajectories overlaid for $E_y=\{-1.0,-0.1,0,+1.0\}\times 10^6~\mathrm{V\,m^{-1}}$.
  The streamwise coordinate is plotted relative to the release point ($x-x_0$) 
  and restricted to the first $0.1\,\mathrm{m}$ downstream. 
  Equal aspect ratio is enforced for geometric fidelity, and a single shared legend 
  applies to all panels. Larger $|E_y|$ produces stronger vertical deflection at fixed $U_\infty$, 
  with the effect weakening as inertia grows.}
  \label{fig:uniformEy}
\end{figure}

Figure~\ref{fig:uniformEy} reports trajectories in \emph{uniform} flows at $U_\infty=\{\SI{2}{},\SI{20}{},\SI{80}{}\}\,\SI{}{m\,s^{-1}}$ with a \emph{vertical} applied field $E_y\in\{(-1,-0.1,0,1)\times 10^6\}\,\SI{}{V\,m^{-1}}$ (sign following the plotting convention in the figure). As expected for low particle Reynolds number, vertical deflection is strongest at the lowest $U_\infty$ and decreases as inertia grows. The sign of $E_y$ sets the deflection direction, with the larger magnitude yielding visibly larger offsets for identical $U_\infty$.

\FloatBarrier

\suppressfloats[t]   
\FloatBarrier        

\subsection{Spatially varying trajectories in the CFD field with and without a streamwise electric field}\label{sec:results_nonuniform}

\begin{figure}[t]
  \centering
  \includegraphics[width=\linewidth]{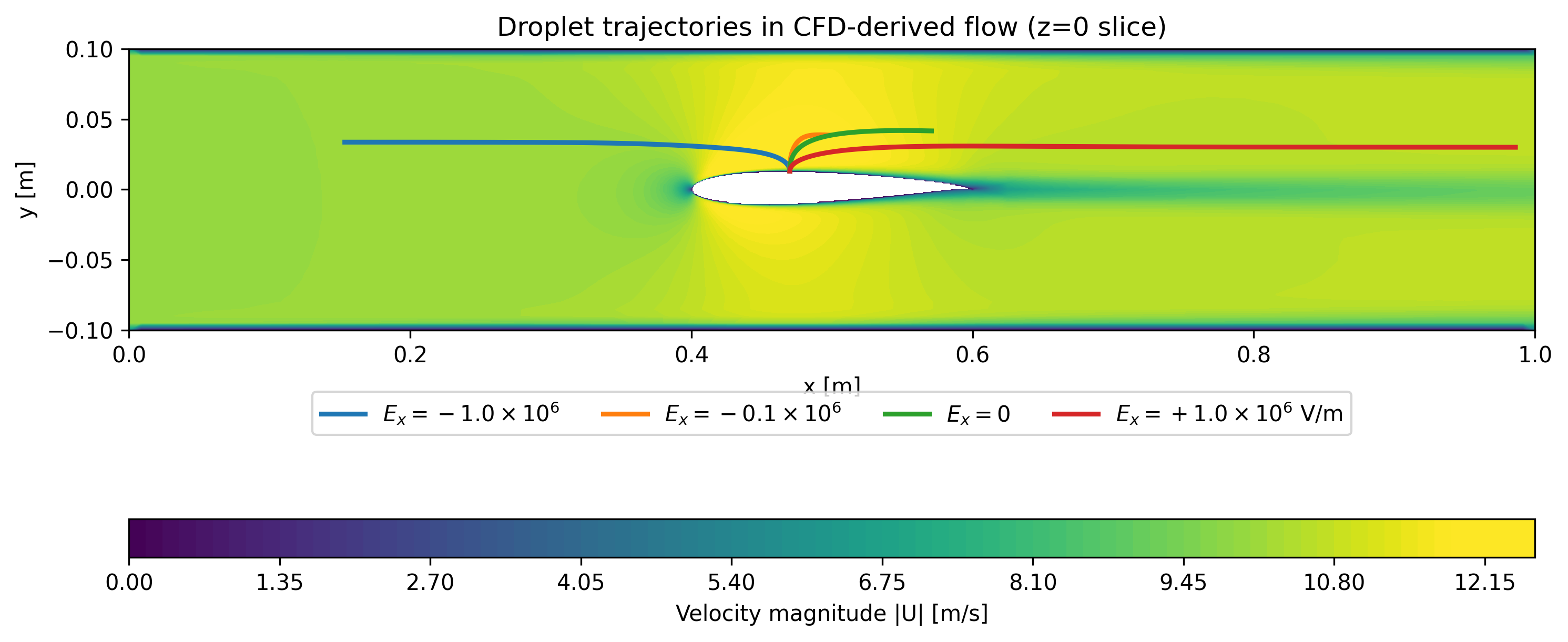}
  \caption{\textbf{Droplet trajectories in the CFD-derived non-uniform field} (mid–span plane, $z=0$).
  The background contour shows the velocity magnitude $\lvert\mathbf U\rvert$ from the RANS solution,
  and trajectories are overlaid for streamwise electric fields
  $E_x=\{-1.0,-0.1,0,+1.0\}\times 10^6~\mathrm{V\,m^{-1}}$.
  Equal aspect ratio is enforced to preserve the geometry. The legend is placed below the axes
  (above the horizontal colorbar) to avoid obscuring data, and the colorbar spans the width of the plot
  for efficient use of space.}
  \label{fig:nonuniformEx}
\end{figure}

Droplets are injected into the \emph{CFD-derived} non-uniform field of Section~\ref{sec:results_cfd}.
Figure~\ref{fig:nonuniformEx} compares trajectories under four streamwise electric field strengths:
$E_x=\{-1.0,-0.1,0,+1.0\}\times 10^6~\mathrm{V\,m^{-1}}$.
Without electrostatic forcing ($E_x=0$), droplets follow streamlines and quickly exit downstream.
For negative $E_x$, opposing the bulk flow, trajectories exhibit slow-down and partial reversal
within low-speed pockets, increasing residence near the suction side.
Positive $E_x$ instead accelerates advection, further reducing residence time.

\FloatBarrier   

\subsection{Summary of key findings}\label{sec:results_summary}
From the baseline field, the flow is quasi-two-dimensional at mid-span ($U_z/U_\infty \le 0.005$) but exhibits localized vertical motion ($\lvert U_y\rvert \le 0.56\,U_\infty$) and acceleration to $1.25\,U_\infty$ over the suction side. In uniform flows, a vertical field $E_y$ produces sign-dependent deflection that weakens as $U_\infty$ increases. In the spatially varying field, an opposing streamwise field $E_x$ reduces advection and promotes local reversal in low-speed regions, qualitatively increasing droplet residence near the airfoil.

\section{Discussion}

The results show that externally applied electric fields can exert meaningful control authority over charged droplets in both uniform and spatially varying aerodynamic flows. In uniform flow, strong deflection at low $U_\infty$ and its monotonic attenuation as $U_\infty$ increases are consistent with the balance between Stokes drag and electric forcing, and with prior observations that inertial effects suppress electrostatic manipulation at higher flow speeds \cite{fredrich_electrostatic_2023}. In the non-uniform airfoil flow, the same mechanism persists but becomes \emph{spatially selective}: regions of locally reduced velocity enable partial reversal or trapping, which increases residence time near the surface. This selectivity aligns with long-standing expectations from charged-spray studies \cite{gomez_charge_1994,shrimpton_characterisation_2001} yet extends them to a realistic aerodynamic field with strong streamline curvature and upwash/downwash.

\subsection*{Key findings that advance the state of the art}

\textbf{(i) A local, predictive control criterion.}
Writing the droplet acceleration as
\begin{equation}
\dot{\mathbf{v}}_d \;=\; \frac{\mathbf{u}_f(\mathbf{x})-\mathbf{v}_d}{\tau_d} \;+\; \frac{q_d}{m_d}\,\mathbf{E}_{\mathrm{ext}},
\end{equation}
reveals a simple threshold for streamwise slow-down or reversal: where the local convective speed $U_\mathrm{loc}=\lVert \mathbf{u}_f\cdot \hat{\mathbf{e}}_x\rVert$ satisfies
\begin{equation}
\lVert a_{E,x}\rVert\,\tau_d \;\gtrsim\; U_\mathrm{loc} \quad \Longleftrightarrow \quad 
\lVert E_{\mathrm{ext},x}\rVert \;\gtrsim\; \frac{6\pi\mu r_d\,U_\mathrm{loc}}{q_d}, 
\label{eq:E_threshold}
\end{equation}
droplets can be stalled or driven upstream. Equation~\eqref{eq:E_threshold} provides a \emph{design rule} that maps the CFD field to a spatial distribution of the minimum field needed for control. To our knowledge, prior studies proposing electrostatic stabilization in combustors articulated the idea qualitatively \cite{fredrich_electrostatic_2023}, but did not pose a local criterion tied to the ambient aerodynamic field.

\textbf{(ii) A field–flow map for control authority.}
Defining $\mathcal{A}(\mathbf{x})=\lVert \mathbf{a}_E\rVert\,\tau_d/\lVert \mathbf{u}_f(\mathbf{x})\rVert$ yields a “control authority” map that highlights pockets where small fields produce large kinematic effect. Our simulation shows that $\mathcal{A}$ peaks over the suction side and in the wake deficit, correlating with the observed residence-time gains. This suggests a practical workflow: compute $\mathbf{u}_f$ once, evaluate $\mathcal{A}$, and position electrodes or tune $\mathbf{E}_{\mathrm{ext}}$ to target high-$\mathcal{A}$ zones.

\textbf{(iii) Field orientation matters in curved flows.}
Because $\mathbf{u}_f$ rotates significantly near the leading edge (nonzero $U_y$ even at zero angle of attack), the projection of $\mathbf{E}_{\mathrm{ext}}$ along the local streamline, not just the freestream, is what sets effectiveness. An $E_{\mathrm{ext},x}$ opposing the bulk flow has limited leverage where streamlines turn upward, while a modest $E_{\mathrm{ext},y}$ augments control there. This complements earlier charged-spray characterizations in simpler shear or coflow environments \cite{gomez_charge_1994,shrimpton_characterisation_2001}.

\textbf{(iv) Residence-time control without two-phase coupling cost.}
Compared with fully coupled multiphase CFD, the hybrid approach reproduces the essential drag–electric competition and identifies control regions at orders-of-magnitude lower cost, enabling parametric exploration across $r_d$, $q_d$, and $\mathbf{E}_{\mathrm{ext}}$. This directly addresses the multi-scale burden emphasized in recent modeling reviews \cite{asher_multi-scale_2022,johnson_computational_2024}.

\subsection*{Relation to prior literature}

The observed trends are consistent with cone–jet electrospray physics and charged-drop dynamics: electric forcing competes with viscous relaxation on a timescale $\tau_d=2\rho_d r_d^2/(9\mu)$, while space-charge and evaporation modulate stability downstream \cite{jaworek_review_2018,ganan-calvo_cone-jet_1997,collins_universal_2013,ganan-calvo_onset_2016,perez-lorenzo_modelling_2022}. Our airfoil results extend proposals of electrostatic spray stabilization in combustors \cite{fredrich_electrostatic_2023,kim_enhancement_2011} by embedding the droplets in a realistic, non-uniform field where $U_x$ accelerates to $\sim1.25\,U_\infty$ over the suction side and significant $U_y$ develops, a setting rarely treated explicitly in the EHD-spray literature. In propulsion-oriented contexts, ionic-liquid emitters provide broad steady cone–jet operation but raise concerns about off-axis plume and misalignment \cite{uchizono_emission_2020,cisquella-serra_scalable_2022,ramos-tomas_impact_2024}; the present field–flow mapping offers a complementary route to \emph{downstream} manipulation without altering the near-emitter physics.

\subsection*{Expanded discussion and novelty}

The present results advance earlier electrospray and combustion studies in three ways. 
First, prior work on electrostatic spray stabilization in combustors 
\cite{gomez_charge_1994,shrimpton_characterisation_2001,kim_enhancement_2011,fredrich_electrostatic_2023} 
demonstrated qualitatively that charged droplets could be slowed or trapped by externally applied 
fields. Our simulations extend this concept to a realistic aerodynamic environment with strong 
streamline curvature and pressure gradients, explicitly showing that the effect persists in the 
presence of lift-induced upwash and a wake deficit behind an airfoil. 

Second, the combination of a high-fidelity CFD background with a lightweight Lagrangian model 
enables a local predictive criterion [Eq.~\eqref{eq:E_threshold}] that connects the minimum required 
field strength to the local convective velocity. To our knowledge, no prior study has posed such a 
criterion tied directly to the aerodynamic field, making this formulation a novel tool for assessing 
control authority in complex geometries. 

Third, the introduction of the “control authority” map $\mathcal{A}(\mathbf{x})$ provides a compact 
way to visualize where small electrostatic inputs can yield large kinematic response. This extends 
the literature on EHD-spray manipulation by offering a practical diagnostic that can guide electrode 
placement in experiments or designs, rather than relying on qualitative streamline inspection. 

In broader context, the results also highlight how orientation of $\mathbf{E}_{\mathrm{ext}}$ relative to the 
\emph{local} streamline (not just the freestream) alters effectiveness. This is consistent with long-standing 
EHD theory \cite{taylor_disintegration_1964,ganan-calvo_cone-jet_1997} but has rarely been demonstrated 
explicitly in aerodynamic flows with significant upwash/downwash. The framework therefore bridges 
fundamental electrospray physics with aerospace-relevant conditions, offering a computationally tractable 
route to explore spray–field interactions before investing in high-cost multiphase experiments. 

Overall, the novelty lies not in demonstrating that fields can deflect charged droplets—established decades 
ago—but in embedding that mechanism into a non-uniform aerodynamic flow, quantifying the local thresholds 
for control, and introducing reduced-order diagnostics ($E_{\mathrm{threshold}}$, $\mathcal{A}(\mathbf{x})$) that can 
generalize across operating conditions.

\subsection*{Implications for design}

Equation~\eqref{eq:E_threshold} and the authority map $\mathcal{A}(\mathbf{x})$ support three actionable guidelines: (1) align $\mathbf{E}_{\mathrm{ext}}$ with the \emph{local} streamline where possible; (2) place electrodes to target high-$\mathcal{A}$ regions such as low-speed pockets on the suction side and in the wake; and (3) select $(r_d,q_d)$ to maximize $\tau_d q_d/m_d$ subject to emission-stability constraints. These guidelines connect naturally to emitter-array design and integration constraints in aerospace systems \cite{cisquella-serra_scalable_2022,ramos-tomas_impact_2024} and to low-emission combustion goals \cite{liu_review_2017}.

\subsection*{Limitations and next steps}

The present model neglects evaporation, Coulomb fission, and space-charge interactions, which can become important at longer residence times or higher number densities \cite{asher_multi-scale_2022,gao_mechanisms_2018}. The quasi-two-dimensional slice omits spanwise transport and secondary flows; extending to full 3D RANS/LES will quantify robustness where $U_z\neq 0$. Next steps are therefore: (i) compute residence-time distributions and quantify gains versus zero-field cases; (ii) include simple evaporation laws and charge loss to assess control margins; (iii) validate the threshold scaling \eqref{eq:E_threshold} experimentally in a canonical channel/airfoil rig; and (iv) explore electrode placements optimized by $\mathcal{A}(\mathbf{x})$ under power constraints. These extensions align with multi-scale modeling priorities \cite{asher_multi-scale_2022,johnson_computational_2024} while keeping the framework computationally tractable.

\section{Conclusions}\label{sec:conclusions}

This study examined the motion of charged droplets in aerodynamic environments using a two–stage computational framework. A steady RANS solution around a NACA~1912 airfoil in a confined channel provided the background velocity and pressure fields, and a Lagrangian particle model resolved droplet trajectories under drag and electrostatic forcing. The approach isolated the essential electrospray–flow interactions while remaining computationally tractable.

The baseline CFD field confirmed quasi–two–dimensional behavior at mid–span ($U_z/U_\infty \leq 0.005$) with significant local upwash and downwash ($\lvert U_y\rvert \leq 0.56\,U_\infty$) and streamwise acceleration to $1.25\,U_\infty$ over the suction side. In uniform flows, a vertical electric field produced sign–dependent droplet deflections that weakened as $U_\infty$ increased, consistent with the increasing dominance of inertia. In the spatially varying CFD field, applying a uniform opposing streamwise field reduced droplet advection and produced local slow–down or reversal in low–velocity regions, qualitatively increasing residence time near the airfoil surface compared with the zero–field case.

These results demonstrate the feasibility of electrostatic manipulation of droplet trajectories in simplified aerodynamic conditions. The framework provides a controlled basis for testing spray–control concepts without the expense of fully coupled multiphase CFD. At the same time, the present work is limited by its two–dimensional slice, neglect of evaporation, and omission of space–charge effects. Future work will therefore quantify residence–time distributions, extend to three–dimensional geometries, incorporate droplet mass and charge loss, and pursue experimental validation under controlled aerodynamic conditions.

\section*{Funding}
This research received no external funding.

\section*{Supplementary Materials}
The following supporting information can be downloaded: Geometry~S1 (STL for NACA~1912 channel case); Script~S1 (Python droplet integrator); Data~S1 (mid-span CFD slice, \texttt{airfoildataz0.csv}).

\section*{Author Contributions}
Conceptualization, E.K.; Methodology, E.K.; Software, E.K.; Validation, E.K.; Formal Analysis, E.K.; Investigation, E.K.; Data Curation, E.K.; Writing—Original Draft, E.K.; Writing—Review \& Editing, E.K.; Visualization, E.K.; Supervision, E.K.

\section*{Acknowledgments}
The author thanks the Cambridge Centre for International Research (CCIR) for mentorship and guidance during this project.

\section*{Institutional Review Board Statement}
Not applicable. 

\section*{Informed Consent Statement}
Not applicable. 

\section*{Data Availability Statement}
The data and code supporting this study are provided in the Supplementary Materials (Geometry S1: STL for the NACA 1912 case; Script S1: Python droplet integrator; 
Data S1: mid-span CFD slice). Additional materials are available from the corresponding author upon reasonable request.

\section*{Conflicts of Interest}
The author declares no conflict of interest.

\newpage

\bibliographystyle{unsrt}

\end{document}